\documentclass[12pt]{article}
\usepackage{epsfig}
\usepackage{psfrag}
\usepackage{enumitem}
\usepackage{latexsym}
\usepackage{indentfirst}
\usepackage{fancyhdr}
\usepackage{amssymb}
\usepackage{amsmath}
\usepackage{amsfonts}
\usepackage{pifont}
\usepackage{cite}
\usepackage{bbold}
\usepackage{ulem}
\usepackage{color}
\usepackage[footnotesize]{caption2}
\usepackage{graphicx}
\usepackage[center,footnotesize,hang]{subfigure}
\usepackage{url}
\usepackage[dvipsnames]{xcolor}
\usepackage{array}
\usepackage{slashed}

\begin{document}
\title{ \hfill\mbox{\small SISSA 54/2016/FISI}\\[-3mm]
\hfill ~\\[0mm]
       \textbf{Lepton Sector Phases and Their Roles in Flavor and Generalized CP Symmetries}
 }
\date{}
\author{\\[1mm]Lisa L.~Everett$^{1\,}$\footnote{E-mail: {\tt
leverett@wisc.edu}}~~and 
Alexander J.~Stuart$^{2,3,4\,}$\footnote{E-mail: {\tt astuart@ucol.mx}}\\
\\[1mm]
  \it{\small $^1$Department of Physics, University of Wisconsin,}\\
  \it{\small Madison, WI 53706, USA}\\[4mm]
 \it{\small $^2$SISSA/INFN, Via Bonomea 265, 34136 Trieste, Italy}\\[4mm]
 \it{\small $^3$Facultad de Ciencias - CUICBAS, Universidad de Colima,}\\ 
\it{\small C.P. 28045, Colima, M\'exico}\\[4mm]
 \it{\small $^4$Dual CP Institute of High Energy Physics, C.P.~28045, Colima, M\'exico}\\ 
 }
\maketitle
\thispagestyle{empty}
\begin{abstract}
\noindent
We study the effects of considering nontrivial unphysical lepton sector phases on the group theoretical properties of the flavor and generalized CP symmetry elements in the case where there are three light, distinct Majorana neutrino species.  We highlight the similarities and differences between the charged lepton and neutrino sectors and further elucidate the group properties of the flavor and generalized CP symmetry elements.  We show how the inclusion of these leptonic phases affects the bottom-up constructions of these symmetry elements and discuss the implications for top-down model building based on discrete symmetry groups.

\end{abstract}

\newpage


\section{Introduction\label{sec:intro}}


With the recent measurements of the reactor mixing angle \cite{dayabay,reno,doublechooz}, the mixing angle pattern of the Maki-Nakagawa-Sakata-Pontecorvo (MNSP) lepton mixing matrix $U_{\text{MNSP}}$~\cite{pdg} is now on solid experimental ground, opening the door to a new era of precision lepton mixing measurements.  As the reactor angle is relatively large (roughly Cabibbo-sized), there is an exciting experimental opportunity for a future direct observation of ``Dirac-type" leptonic CP violation.  Even in the absence of such a direct measurement, hints of a nonzero value of the leptonic Dirac CP-violating phase $\delta$ are already originating in global fit analyses of lepton mixing data \cite{global}.   Thus in preparation for a nonzero measurement of this phase,  it is useful to explore whether the underlying theory violates CP explicitly,  or if CP is a spontaneously broken symmetry.

In the case in which CP is conserved in the underlying theory and is broken spontaneously,  it is standard to explore this question in the context of theories with spontaneously broken \textit{discrete} lepton family symmetries (see \cite{kingluhnreview} for a review). For the case of Majorana neutrinos, one compelling idea within this framework is to assume the breaking of a high energy discrete flavor symmetry to a residual Klein symmetry in the neutrino sector.  The residual Klein symmetry completely fixes the form of $U_{\text{MNSP}}$ at leading order, in the diagonal charged lepton basis up to charged lepton rephasing (although it fails to predict values for the Majorana phases\cite{majorana1,majorana2}). 
To make predictions for the Majorana phases, a CP symmetry can be imposed and then spontaneously broken, resulting in concrete phase predictions.  The CP symmetry should of course be defined consistently together with the discrete flavor symmetry\cite{lindner}.  Many such models of CP and flavor symmetries have been proposed and investigated, including models based on $A_4$\cite{A4CP1,A4CP2}/$\Delta(3n^2)$\cite{Nishi:2013jqa,Ding:2014hva,Hagedorn:2014wha,Ding:2015rwa}, $A_5$\cite{A5CP1,A5CP2,A5CP3,A5CP4},  $S_4$\cite{Feruglio:2012cw,S4CP,Feruglio:2013hia,Li:2013jya,bimaxS4,S4Z2CP}/$\Delta(6n^2)$\cite{Hagedorn:2014wha,NandK,Ding:2014ssa,Ding:2014ora,Chen:2015nha}, $\Sigma(n\phi)$\cite{sigmagroups,Rong:2016cpk}, $T'$\cite{tprime}, and  $D^{(1)}_{9n, 3n}$\cite{D9n3n}.\footnote{We note that CP has been studied for the case of a single preserved $Z_2$ as the residual neutrino flavor symmetry\cite{Luhn:2013lkn,Z2CP1,Z2CP2,Z2CP3,Z2CP4,S4Z2CP}.}   

With these studies in mind, is also worthwhile to take a bottom-up perspective, in which the goal is not to construct a specific top-down model, but instead to elucidate how and when preserving different residual CP and flavor symmetry elements affects the predictions for the lepton mixing parameters. To this end, in previous work \cite{bottomup} we constructed the general residual Klein and generalized CP symmetries in the neutrino sector as a function of the measured lepton mixing parameters.  There the simplifying assumption was made that the charged lepton sector is diagonal, such that the neutrino sector mixing parameters are directly related to the experimentally measured (measurable) lepton mixing parameters.

 In this paper, we consider the role of general leptonic sector phases.  These phases include charged lepton sector phases and overall shifts to Majorana phases, which are typically ignored since by definition they cannot have any effect on physical observables.  However, their inclusion clarifies the group theoretical properties at low energies of several quantities of interest for theoretical model-building, which can be obscured when these phases are set to zero.  In particular, these phases have relevance for making connections of the family and generalized CP symmetry elements to explicit representations of specific discrete symmetry groups, and hence their inclusion provides a natural generalization of the bottom-up approach given in \cite{bottomup}.  Our results provide a set of group theoretical relations that must be satisfied at low energies within any top-down flavor model-building scenario for three light Majorana neutrinos that leaves a residual Klein symmetry in the neutrino sector.  The goal is to provide guidelines for developing a better understanding of the generalized CP and flavor symmetries when constructing top-down models, which can in principle lead to new model-building directions within this general framework.

This paper is structured in the following way.  In Section~\ref{sec:frameworkflav}, we examine the effects that phase redefinitions in lepton mixing can have on the group structures of the underlying flavor symmetries.  We will discuss the effects of elevating the status of certain subsets of these phases to that of flavor symmetries can have on the form of $U_{\text{MNSP}}$, and analyze the group structures of such choices of rephasings/symmetries with a special focus on discrete symmetry groups. In Section~\ref{sec:frameworkCP}, we expand this discussion to include the ways in which such phases affect aspects of generalized CP symmetries.  Section~\ref{sec:examples} provides a detailed exposition of the connections of this work to the bottom-up approach given in \cite{bottomup}, including several examples.  We present our conclusions and outlook in Section~\ref{sec:conclusion}.


\section{Phases and Flavor Symmetries\label{sec:frameworkflav}}






The starting point of this analysis is the Majorana neutrino mass matrix for the three light neutrino species, $M_{\nu}$. It is diagonalized by the matrix $U_{\nu}$, as follows:
\begin{equation}\label{eq:diagMnu}
U_{\nu}^TM_{\nu}U_{\nu}=M_{\nu}^{\text{Diag}}=\text{Diag}(m_1,m_2,m_3)=\text{Diag}(\vert m_1\vert e^{-i\alpha_1},\vert m_2\vert e^{-i\alpha_2},\vert m_3\vert e^{-i\alpha_3}),
\end{equation}
in which $\vert m_{1,2,3}\vert$ are presumed to be nondegenerate and nonzero, i.e., $\vert m_1\vert \neq \vert m_2\vert \neq \vert m_3\vert \neq 0$.  The transformation
\begin{equation}\label{eq:baseunphysnu}
U_{\nu}\rightarrow U_{\nu}Q_{\nu}, \text{~with~} Q_{\nu}=\text{Diag}((-1)^{p_1}, (-1)^{p_2}, (- 1)^{p_3}) \text{~where~$p_{1,2,3}=0,1$}
\end{equation}
also diagonalizes $M_{\nu}$ and leaves $M_{\nu}^{\text{Diag}}$ invariant.\footnote{It is possible to put an overall phase in $U_{\nu}$, i.e., $U_{\nu}\rightarrow U_{\nu}e^{i\theta_{\nu}/2}$, without changing any physical predictions.  This is a point that we will revisit later.}  There are thus eight possible symmetries contained in $Q_\nu$, corresponding to the eight possible assignments of $p_{1,2,3}$ as given above.

In the charged lepton sector, the mass matrix $M_e=m_em_e^{\dagger}$, which connects left-handed states, is diagonalized by $U_e$, as follows:
\begin{equation}\label{eq:diagMe}
U_{e}^{\dagger}M_{e}U_{e}=M_{e}^{\text{Diag}}=\text{Diag}(|m_e|^2,|m_{\mu}|^2,|m_{\tau}|^2),
\end{equation}
in which again $|m_e|\neq|m_{\mu}|\neq|m_{\tau}|\neq 0$.
The diagonalization of $M_e$ thus can easily be seen to be left invariant by the transformation
\begin{equation}\label{eq:baseunphyscl}
U_e\rightarrow U_e Q_e,\text{ where }Q_e=\text{Diag}(e^{i\beta_1},e^{i\beta_2},e^{i\beta_3}),\text{~where $\beta_{1,2,3}\in [0,2\pi)$}.
\end{equation}
Eqs.~\eqref{eq:baseunphysnu} and \eqref{eq:baseunphyscl}  represent the set of transformations which leave the (diagonal) mass matrices of Eqs.~\eqref{eq:diagMnu} and \eqref{eq:diagMe} invariant. 
As these transformations play no role in the diagonalization of $M_{\nu}$ ($M_e$), they cannot enter any physical predictions that arise from $U_{\nu}$ ($U_e$). More precisely, $U_{\nu}\rightarrow U_{\nu}Q_{\nu}$ and $U_e\rightarrow U_eQ_e$ imply that the MNSP matrix correspondingly transforms as
\begin{equation}\label{eq:MNSPinv}
U_{\text{MNSP}}= U_e^\dagger U_\nu \rightarrow   Q_e^{\dagger}U_{\text{MNSP}}Q_{\nu}=U_{\text{MNSP}}',
\end{equation}
and that $U_{\text{MNSP}}$ and (the infinitely many possible) $U_{\text{MNSP}}'$ must all yield the same physics predictions (see~\cite{rephasinginv} for a similar discussion in terms of rephasing invariants).  The utility of including unphysical phases in Eq.~\eqref{eq:MNSPinv} can be seen by observing  that  $Q_e$ and $Q_{\nu}$ can be related to their nondiagonal forms $T_e$ and $S_{\nu}$ via the unitary transformations \cite{bottomup}
\begin{equation}\label{eq:symdtond}
S_{\nu}=U_{\nu}Q_{\nu}U_{\nu}^{\dagger}, ~T_e=U_eQ_eU_e^{\dagger},
\end{equation}
 These relationships can be  derived by using  Eq.~\eqref{eq:diagMnu} and Eq.~\eqref{eq:diagMe} as well as the condition for a flavor symmetry in its nondiagonal basis,  i.e.,
\begin{equation}\label{eq:flasymnondiag}
S_{\nu}^TM_{\nu}S_{\nu}=M_{\nu},~T_e^{\dagger}M_eT_e=M_e.
\end{equation}

Up to this point, we have not yet specified the explicit forms of the mixing matrices $U_e$ and $U_{\nu}$.  Within the top-down approach, one constructs a concrete model of $M_e$ and $M_\nu$ in a specific flavor basis (for example, it is often taken to be the basis in which $M_e$ is diagonal).  $U_e$ and $U_\nu$ are then found through explicit diagonalization (up to rephasings).   By contrast, from a bottom-up perspective, the forms of $U_e$ and $U_\nu$ can be fixed or arbitrary depending on the choice of and/or number of symmetry elements that are to imposed from Eq.~\eqref{eq:baseunphysnu} and Eq.~\eqref{eq:baseunphyscl}.   For example, in \cite{bottomup} we explicitly constructed the forms of the $S_{\nu}$ in the case in which the four positive determinant choices of $Q_\nu$ are preserved, while the charged leptons are taken to be diagonal. In this case, the mixing parameters of $U_\nu$ have a direct connection to the measured mixing parameters of $U_{\rm MNSP}$, and thus the flavor symmetry elements can be given explicitly in terms of measurable quantities\cite{bottomup}.   Here we will explore more general situations in which we allow for different possibilities for the choice and number of conserved symmetry elements.  The upshot of this discussion is that although the unphysical phases contained in $Q_e$ and $Q_{\nu}$ do not enter any physical observables by construction, they can clearly play a critical role in model building when it pertains to fixing/predicting lepton mixing patterns.

To see the way in which imposing specific symmetry elements can fix the mixing parameters, it is worthwhile first to consider the case in which there is a two-fold degeneracy in the entries of $Q_\nu$ and $Q_e$.   Applying Eq.~\eqref{eq:symdtond}   to Eq.~\eqref{eq:flasymnondiag} shows that if  $p_i=p_j$ for some $i,j=1,2,3$ or $\beta_k=\beta_l$ for some $k,l=1,2,3$, cf.~Eq.~\eqref{eq:baseunphysnu} or \eqref{eq:baseunphyscl} respectively, then there will exist an additional unitary rotation $U^{\nu}_{ij}$ or $U^e_{kl}$ that allows for the mixing of the degenerate states, such that
\begin{equation}\label{eq:2phasedegen}
Q_{\nu}^{T}(U^{\nu}_{ij})^{T}M_{\nu}^{\text{Diag}}U^{\nu}_{ij}Q_{\nu}=(U^{\nu}_{ij})^{T}M_{\nu}^{\text{Diag}}U^{\nu}_{ij},
~Q_e^{\dagger}(U^e_{kl})^{\dagger}M_e^{\text{Diag}}U^e_{kl}Q_e=(U^e_{kl})^{\dagger}M_e^{\text{Diag}}U^e_{kl},
\end{equation}
in which $(Q_\nu)_{ii}=(Q_\nu)_{jj}$ and $(Q_e)_{kk}=(Q_e)_{ll}$.  This shows that when there are two degenerate phases in either  $Q_e$ or $Q_{\nu}$, this is not enough to fix $M_e$ or $M_{\nu}$ to be of diagonal form.  Said again, a two-fold degeneracy in either $Q_e$ or $Q_{\nu}$ is a symmetry not only of $M_e^{\text{Diag}}$ or $M_{\nu}^{\text{Diag}}$, but more importantly, $(U^e_{kl})^{\dagger}M_e^{\text{Diag}}U^e_{kl}$ and $(U^{\nu}_{ij})^{T}M_{\nu}^{\text{Diag}}U^{\nu}_{ij}$, cf.~Eq.~\eqref{eq:2phasedegen}.  Thus, to diagonalize these mass matrices completely, it is necessary to map
\begin{equation}\label{eq:1paramdiag}
U_{\nu}\rightarrow U_{\nu}U^{\nu}_{ij}Q_{\nu}\text{ and } U_e\rightarrow U_e U^e_{kl}Q_e,
\end{equation}
which implies that the mixing matrices $U_{\nu}$ and $U_e$ are only fixed up to these unitary rotations in the degenerate sub-block of eigenvalues.\footnote{A similar situation occurs if all phases are equal in either $Q_e$ or $Q_{\nu}$, i.e.,~$p_i=p_j=p_k$ and $\beta_i=\beta_j=\beta_k$ in Eqs.~\eqref{eq:baseunphysnu} and \eqref{eq:baseunphyscl} respectively.  In this case, the mixing is fixed up to an arbitrary unitary $3\times 3$ rotation.} The explicit forms of these extra unitary rotations are of course specified by diagonalizing $M_e$ and $M_{\nu}$.

The preceding discussion assumed  a single choice of phases for $Q_e$ and $Q_{\nu}$.  However, if we impose additional choices for the phases, then this can  in principle fix the mixing.  For example, let us consider the neutrino sector, for which there are eight possible symmetries contained in/denoted by $Q_{\nu}$.  Further imposing $Q_{\nu}=\pm 1$ does not change the result; neither does $Q_{\nu}'=-Q_{\nu}$, i.e., the other phase assignment with the same degenerate sub-block.  What clearly affects the result is to impose an additional phase choice such that  $(Q_\nu)_{kk}=(Q_\nu)_{ll}$ for some $k,l$ yet to be determined.  By demanding that this phase choice also holds, we obtain
\begin{equation}\label{eq:2symsnu}
(U^{\nu}_{ij})^{T}M_{\nu}^{\text{Diag}}U^{\nu}_{ij}=(U^{\nu}_{kl})^{T}M_{\nu}^{\text{Diag}}U^{\nu}_{kl}\implies
M_{\nu}^{\text{Diag}}=(U^{\nu}_{ij}(U_{kl}^{\nu})^{\dagger})^{T}M_{\nu}^{\text{Diag}}U^{\nu}_{ij}(U_{kl}^{\nu})^{\dagger}.
\end{equation}
If $kl=ij$ no additional constraints arise, and the mixing still contains the same number of free parameters (after angle and phase redefinitions).  However, if $ij\neq kl$, a comparison of  Eq.~\eqref{eq:2symsnu} with Eqs.~\eqref{eq:diagMnu}-\eqref{eq:baseunphysnu} demonstrates that the arbitrary rotation angles and phases in  $U_{ij}^{\nu}$ and $U_{kl}^{\nu}$ are now related to each other because in this case
\begin{equation}\label{eq:UUtoQnu}
U^{\nu}_{ij}(U_{kl}^{\nu})^{\dagger}=\text{Diag}(\pm 1, \pm 1, \pm 1)=Q_{\nu}.
\end{equation}
Hence, the transformation $U_{\nu}\rightarrow U_{\nu}U^{\nu}_{ij}Q_{\nu}$ in Eq.~\eqref{eq:1paramdiag} is then reduced  to $U_{\nu}\rightarrow U_{\nu}Q_{\nu}$, eliminating the additional mixing parameters that entered when only preserving a single set of phases containing a two-fold degeneracy.

This discussion can be generalized to the charged lepton sector, with one notable difference that stems from the fact that the $Q_\nu$ contain only $\pm 1$, while the $Q_e$ depend on the arbitrary $\beta_i$ phases. We have seen that for the neutrino sector, two such nontrivial rephasings were required to be chosen subject to Eqs.~\eqref{eq:2symsnu}-\eqref{eq:UUtoQnu}, in order to guarantee the mixing matrix $U_{\nu}$ has no additional free parameters that result from a degenerate sub-block.   However, there is clearly no such constraint for $Q_e$ because the phases contained in $Q_e$ can all be chosen to be distinct, as seen in Eq.~\eqref{eq:baseunphyscl}. This then  fixes the mixing matrix $U_e$,  and thus it fixes the mass matrix $M_e$, up to rephasings by $Q_e$.




Let us now consider the case in which such rephasings are obtained from a flavor symmetry group, for which these arguments can be described within a group theoretical framework.   To this end, we note that in the preceding discussion, the forms of $U_e$ and $U_{\nu}$ were completely fixed up to rephasing by $Q_e$ and $Q_{\nu}$ (with the possibility of an additional rotation when there is a degeneracy of phases in $Q_e$ or $Q_{\nu}$, cf.~Eq.~\eqref{eq:1paramdiag}).  Perhaps just as importantly,   Eq.~\eqref{eq:symdtond} demonstrates that  $U_{\nu}$ and $U_{e}$ are the unitary transformations that relate $Q_{\nu}$ to $S_{\nu}$ and $Q_e$ to $T_e$.  As such, $Q_e$ and $Q_{\nu}$ can be interpreted as representations of the elements of a flavor symmetry in their corresponding diagonal bases.  Therefore to understand \textit{all} possible residual flavor symmetries, we need only to understand the group properties of $Q_e$ and $Q_{\nu}$.  

We start by noting that the unphysical phases of Eqs.~\eqref{eq:baseunphysnu} and \eqref{eq:baseunphyscl} generally take multiple values, i.e., $Q_{\nu}$ represents a collection of eight symmetry transformations and $Q_e$ represents an infinite set of symmetry transformations parameterized by the continuous parameters $\beta_{1,2,3}$.  Therefore, the full residual neutrino flavor symmetry group $G_{\nu}$ and the full residual charged lepton flavor symmetry group $G_e$ are expressible as\footnote{We emphasize again that these symmetry groups are contingent upon having nonzero, nondegenerate lepton masses.}
\begin{equation}\label{eq:flavsymfull}
G_{\nu}\cong Z_2^{p_1}\times Z_2^{p_2} \times Z_2^{p_3},~G_e\cong U(1)_{\beta_1}\times U(1)_{\beta_2} \times U(1)_{\beta_3},
\end{equation}
(recall~Eqs.~\eqref{eq:baseunphysnu} and \eqref{eq:baseunphyscl}).  However, the previous discussion shows that it is not necessary to implement the totality of these symmetries to generate a specific mixing pattern (although doing so certainly will).  Thus with an eye toward minimality,  we observe that it is possible to rewrite Eq.~\eqref{eq:diagMe} as
\begin{equation}\label{eq:diagMephase}
M_e^{\text{Diag}}=U_e^{\dagger}M_eU_e=U_e^{\dagger}(P_e^{\dagger}P_e)M_e(P_e^{\dagger}P_e)U_e=U_e^{\dagger}P_e^{\dagger}M_eP_eU_e,
\end{equation}
in which $P_e=e^{i\theta_e}$.  Clearly, $U_e\rightarrow P_eU_eQ_e$ still diagonalizes $M_e$. The phase in $P_e$ can always be chosen to fix $\text{Det}(Q_e)=+1$, for example by choosing $\theta_e=-(\beta_1+\beta_2+\beta_3)/3$, cf.~Eq.~\eqref{eq:baseunphyscl}.

The same result can be obtained for the neutrino sector, i.e., $\text{Det}(Q_{\nu})=+1$, but for different reasons.  More precisely, since we have assumed neutrinos are Majorana fermions, the freedom does not exist to rephase the whole mass matrix $M_{\nu}$ by arbitrary phases without affecting the (complex) neutrino mass eigenvalues.  However, global phases on Majorana mass matrices are phenomenologically irrelevant because they contribute to an overall shift of each of the individual phases (which is unmeasurable), i.e.,
\begin{equation}\label{eq:diagMnuphase}
M_{\nu}^{\text{Diag}}=U_{\nu}^TM_{\nu}U_{\nu}\rightarrow (P'_{\nu})^TM_{\nu}^{\text{Diag}}P'_{\nu}=(P'_{\nu})^TU_{\nu}^TM_{\nu}U_{\nu}P'_{\nu}=U_{\nu}^T(P'_{\nu})^TM_{\nu}P'_{\nu}U_{\nu},
\end{equation}
where $P'_{\nu}=e^{-i\theta_{\nu}'/2}$.  Then, letting $\theta_{\nu}'=\theta_{\nu}+p\pi$ ($p$ an integer) implies that
\begin{equation}\label{eq:decompP'nu}
P_{\nu}'=(-1)^p P_{\nu},
\end{equation}
in which $P_{\nu}=e^{-i\theta_{\nu}/2}$.  This allows for the shift $\alpha_{1,2,3}\rightarrow\alpha_{1,2,3}'=\alpha_{1,2,3}+\theta_{\nu}$, cf.~Eq.~\eqref{eq:diagMnu}, and the determination of $\text{Det}(Q_{\nu})=+1$ by utilizing the $(-1)^p$ factor in Eq.~\eqref{eq:decompP'nu}, with $p=p_1+p_2+p_3$, cf.~Eq.~\eqref{eq:baseunphysnu}.\footnote{Compare to $\theta_e=-(\beta_1+\beta_2+\beta_3)/3$.}    To summarize, it is possible to restrict $\text{Det}(Q_{\nu})=\text{Det}(Q_e)=+1$ to remove physically redundant symmetries so that the elements of the minimal, residual leptonic symmetries $G_\nu$ and $G_e$ can be expressed as\footnote{This is a slight abuse of notation for $p_{1,2,3}$ and $\beta_{1,2,3}$, as they \textit{actually} should be $p'_{1,2,3}$ and $\beta_{1,2,3}'$, where $p_1'=2p_1+p_2+p_3$, $p_2'=2p_2+p_1+p_3$, $p_3'=2p_3+p_1+p_2$, $3\beta'_1=2\beta_1-\beta_2-\beta_3$, $3\beta'_2=2\beta_2-\beta_1-\beta_3$, $3\beta'_3=2\beta_3-\beta_1-\beta_2$. }
\begin{equation}\label{eq:flavsymdet1}
G_{\nu}=\text{Diag}((-1)^{p_2+p_3},(-1)^{p_2},(-1)^{p_3}),~G_e=\text{Diag}(e^{-i(\beta_2+\beta_3)},e^{i\beta_2},e^{i\beta_3}),
\end{equation}
in which $p_{2,3}=0,1$ and $\beta_{2,3}\in [0,2\pi)$. This restriction is equivalent to ``removing'' the four $Q_{\nu}$ with $\text{Det}(Q_{\nu})=-1$ and the infinitely many $Q_e$ with $\text{Det}(Q_e)=e^{i\theta_e}$, where $\theta_e\in(0,2\pi)$.

Motivated by these ``new'' general forms for $G_{\nu}$ and $G_e$, we next define
\begin{equation}\label{eq:flavsymdecomp}
\begin{aligned}
\widetilde{G}_{p_2}=\text{Diag}((-1)^{p_2},(-1)^{p_2},1),&~ \widetilde{G}_{p_3}=\text{Diag}((-1)^{p_3},1,(-1)^{p_3}), \\
\widetilde{T}_{\beta_2}=\text{Diag}(e^{-i\beta_2},e^{i\beta_2},1),& ~\widetilde{T}_{\beta_3}=\text{Diag}(e^{-i\beta_3},1,e^{i\beta_3}),
\end{aligned}
\end{equation}
for all $p_{2,3}=0,1$ and for all $\beta_{2,3}\in [0, 2\pi)$.  From these definitions, we see that the mappings 
\begin{equation}\label{eq:z2z2iso}
\widetilde{G}_{p_2}\rightarrow (-1)^{p_2}\text{ and }\widetilde{G}_{p_3}\rightarrow (-1)^{p_3}
\end{equation}
for all $p_{2,3}=0,1$, imply $\widetilde{G}_{p_2}\cong Z_2^{p_2}$ and  $\widetilde{G}_{p_3}\cong Z_2^{p_3}$.
Furthermore, it is clear that 
$[\widetilde{G}_{p_2},~\widetilde{G}_{p_3}]=0$ 
for every $p_{2,3}=0,1$.  Hence, $G_{\nu}\cong Z_2^{p_2}\times Z_2^{p_3}$.   

A similar situation occurs  in the charged lepton sector, in which it is possible to uniquely map
\begin{equation}\label{eq:u1u1iso}
\widetilde{T}_{\beta_2}\rightarrow e^{i\beta_2}\text{ and } \widetilde{T}_{\beta_3}\rightarrow e^{i\beta_3}
\end{equation}
for all $\beta_{2,3}\in [0,2\pi)$, implying $\widetilde{T}_{\beta_2}\cong U(1)_{\beta_2}$ and  $\widetilde{T}_{\beta_3}\cong U(1)_{\beta_3}$.  The $\widetilde{T}_{\beta_{2,3}}$ also commute ($[\widetilde{T}_{\beta_2},~\widetilde{T}_{\beta_3}]=0$ for all $\beta_{2,3}\in [0,2\pi)$).
Therefore, $G_e\cong U(1)_{\beta_2}\times U(1)_{\beta_3}$.   Hence, the most general, non-redundant set of residual lepton symmetries is
\begin{equation}\label{eq:flavsymrestrict}
G_{\nu}\cong Z_2^{p_2}\times Z_2^{p_3},~G_e\cong  U(1)_{\beta_2}\times U(1)_{\beta_3},
\end{equation}
in which we remind the reader that the superscripts (subscripts) parameterize all elements of the discrete (continuous) symmetry group $G_{\nu}~(G_e)$.

Within the context of discrete lepton family symmetries, it is clear that to obtain both $G_{\nu}$ and $G_e$ from the spontaneous breaking of a specific discrete family  symmetry group $G_f$, the infinite parameter space of the continuous group $G_e$ must be restricted.  To this end,  let us consider the situation in which $\beta_{2,3}=2\pi k_{2,3}/n_{2,3}$, where  $n_{2,3}$ are integers that divide the order of the discrete group $G_f$ and $k_{2,3}=0,\ldots,n_{2,3}-1$.  
In this case, the set of residual charged lepton symmetries is expressible as
\begin{equation}\label{eq:diagGe2gen}
T^{\text{Diag}}_{k_2/n_2, k_3/n_3}=\text{Diag}( e^{ -2\pi i (n_3 k_2+n_2 k_3)/n_2n_3}, e^{ 2\pi i k_2/n_2}, e^{2\pi i k_3/n_3}),
\end{equation}
in which $k_{2,3}=0,\ldots,n_{2,3}-1$.  This clearly represents the elements of $G_e\cong Z_{n_2}\times Z_{n_3}$.  Furthermore, if the mixing from the charged lepton sector is to be completely determined, this  fixes $n_{2,3}\geq 2$ from the previous logic when discussing two-fold rephasing degeneracies.

Eq.~\eqref{eq:diagGe2gen} demonstrates that it is never possible to have a diagonal entry of $T^{\text{Diag}}_{k_2/n_2,k_3/n_3}$ that is always real (i.e.,~it is unity), because $k_{2,3}/n_{2,3}=0$ for only $k_{2,3}=0$ and not the remaining $k_{2,3}=1,\ldots n_{2,3}-1$. To impose this criteria, one of the phases, for example~$\beta_{1,2,3}$ in Eq.~\eqref{eq:flavsymdet1}, must be set to zero.  Without loss of generality, let us take $\beta_2=-\beta_3=\beta$ ($k_2/n_2=-k_3/n_3=k/n$), so that $\beta_1=0$.  Equation~\eqref{eq:diagGe2gen} then becomes
\begin{equation}\label{eq:diagGe1gen}
T^{\text{Diag}}_{k/n}=\text{Diag}( 1, e^{- 2\pi i k/n}, e^{2\pi i k/n}),
\end{equation}
in which $k=0,\ldots, n-1$.  Physically, this case corresponds to the complete breaking of one of the original $U(1)$ symmetries, because $Z_n\cong Z_1\times Z_n$, in which $Z_1$ is the (trivial) symmetry of a completely broken $U(1)$. Then, as previously discussed, to completely fix charged lepton mixing in this case, it is necessary to impose  $n\geq 3$.  Therefore, for the remainder of this work, we will assume that  if $G_e\cong Z_{n_2}\times Z_{n_3}$, then $n_{2,3}\geq 2$, or if $G_e\cong Z_n$, then $n\geq 3$.

Turning to the neutrino sector, we see that removing the redundant rephasing choices from Eq.~\eqref{eq:baseunphysnu} with $\text{Det}(Q_{\nu})=-1$ yields the well-known result that in the diagonal basis, the elements $(G_\nu)_{i=0,1,2,3} \equiv G_i^\text{Diag}$ are given by
\begin{equation}\label{eq:diagGs}
\begin{aligned}
&G^{\text{Diag}}_0=\text{Diag}(1,1,1),~G_1^{\text{Diag}}=\text{Diag}(1,-1,-1),\\&G_2^{\text{Diag}}=\text{Diag}(-1,1,-1),~G_3^{\text{Diag}}=\text{Diag}(-1,-1,1).
\end{aligned}
\end{equation}
This is of course the aforementioned group  $G_{\nu}\cong Z_2^{p_2}\times Z_2^{p_3}$ (cf.~Eq.~\eqref{eq:flavsymrestrict}) that is isomorphic to the Klein symmetry group.  The previous logic then dictates that we must preserve/impose two\footnote{Actually, when preserving two nontrivial elements of such a symmetry, the third comes for free because $G^{\text{Diag}}_iG_j^{\text{Diag}}=G^{\text{Diag}}_k$ for $i\neq j\neq k$.}  of these nontrivial elements to completely fix $U_\nu$ because preserving only one nontrivial element yields a  ``free'' unitary rotation (see Eqs.~\eqref{eq:2phasedegen}-\eqref{eq:1paramdiag}).




An inspection of Eq.~\eqref{eq:diagGs}  demonstrates another striking (but perhaps obvious) difference between the largest possible sets of non-redundant neutrino symmetries and charged lepton symmetries.  The issue is that while the order of all of the nontrivial neutrino symmetry elements is two, the order of the nontrivial charged lepton symmetries fluctuates depending on the ratio of $\beta_{2,3}=k_{2,3}/n_{2,3}$ or $\beta=k/n$.   More precisely, Eqs.~\eqref{eq:diagGe2gen}-\eqref{eq:diagGe1gen} show that the order of the elements of $G_e$ can oscillate between multiple values depending on the value of $k_{2,3}=0,1,\ldots,n_{2,3}-1$ or $k=0,\ldots , n-1$.    Hence, we see that
\begin{equation}\label{eq:orderGe}
(T^{\text{Diag}}_{k_2/n_2, k_3/n_3})^{n_2n_3}=1,~ (T^{\text{Diag}}_{k/n})^n=1.
\end{equation}
The orders of $T^{\text{Diag}}_{k_2/n_2, k_3/n_3}$ and $T^{\text{Diag}}_{k/n}$ are thus given by $n_2n_3$ and $n$, respectively, unless special conditions are met.  For example, if there exist integers $l_{2,3}$ and $l$ such that $n_{2,3}=l_{2,3}k_{2,3}$ and $n=lk$,\footnote{Note that $l_{2,3},l\neq 1$ because by definition $k=0,\ldots, n-1$ and $k_{2,3}=0,\ldots, n_{2,3}-1$.   } then
\begin{equation}\label{eq:orderGered}
(T^{\text{Diag}}_{k_2/n_2, k_3/n_3})^{l_2l_3}=(T^{\text{Diag}}_{1/l_2,1/ l_3})^{l_2l_3}=1,~ (T^{\text{Diag}}_{k/n})^l= (T^{\text{Diag}}_{1/l})^l=1,
\end{equation}
which reduces the order of  $T^{\text{Diag}}_{k_2/n_2, k_3/n_3}$ from $n_2n_3$ to $l_2l_3$ and the order of $T^{\text{Diag}}_{k/n}$ from $n$ to $l$.  A further reduction occurs in $T^{\text{Diag}}_{1/l}$ if $l=0$ ($k=0$), as  the order of $T^{\text{Diag}}_{l}$ then is $1$.  Similarly, if  $k_2$ ($k_3$) is $0$,  the order of $T^{\text{Diag}}_{k_2/n_2, k_3/n_3}$ is $n_3$ ($n_2$), and if $k_2=k_3=0$, the order of  $T^{\text{Diag}}_{0,0}$ is $1$.

Hence, an investigation of the orders of all possible elements that can be contained in the residual symmetries $G_e$ and $G_{\nu}$ (assuming these symmetries originate from a discrete symmetry group $G_f$), shows different results for the two sectors.  More precisely, $G_{\nu}$ contains only (3) order 2, nontrivial elements, whereas the orders of the nontrivial elements of $G_e$ can have a wide range of values. These values depend on the ratios $\beta_{2,3}=k_{2,3}/n_{2,3}$ ($\beta=k/n$), in which $k_{2,3}=0,\ldots, n_{2,3}-1$, and $n_{2,3}\geq 2$ to completely fix charged lepton mixing.


\section{Phases and Generalized CP\label{sec:frameworkCP}}





Having analyzed the relevance of unphysical phases in fixing lepton mixing predictions and explored their possible group theoretical origin in terms of flavor symmetries, we next expand the logic as set forth in Section \ref{sec:frameworkflav} to include that of generalized CP symmetries.  
 These conditions take the form\cite{WBinv1}
\begin{equation}\label{eq:defXnu}
X_{\nu}^TM_{\nu}X_{\nu}=M_{\nu}^*
\end{equation}
and
\begin{equation}\label{eq:defYe}
Y_e^{\dagger}M_eY_e=M_e^*.
\end{equation}
In analogy  to the the case of the flavor symmetries $T_e$ and $S_{\nu}$ as defined in Eq.~\eqref{eq:flasymnondiag}, $Y_e$ and $X_{\nu}$ represent (possibly infinite) sets of transformations. 

To understand the similarities and differences between Eqs.~\eqref{eq:defXnu}--\eqref{eq:defYe}, we will elaborate on the discussion in~\cite{bottomup}, so that it may be easily extended/connected to the charged lepton sector.  We begin with the diagonal neutrino basis.  From~Eq.~\eqref{eq:defXnu}, we have
\begin{equation}\label{eq:diagnuCP}
(X_{\nu}^{\text{Diag}})^TM_{\nu}^{\text{Diag}}X_{\nu}^{\text{Diag}}=(M_{\nu}^{\text{Diag}})^*
\end{equation}
with
\begin{equation}\label{eq:Xdiag}
X_{\nu}^{\text{Diag}}=\text{Diag}(\pm e^{i\alpha_1'},\pm e^{i\alpha_2'},\pm e^{i\alpha_3'}).
\end{equation}
 In the above, $\alpha'_i$ is given by $\alpha'_i=\alpha_i+\theta_{\nu}$, in which $\theta_{\nu}$ keeps track of an arbitrary global phase redefinition of $M_{\nu}$.     It is next useful to define
\begin{equation}\label{eq:X0}
X_0^{\text{Diag}}=\text{Diag}(e^{i\alpha_1'},e^{i\alpha_2'},e^{i\alpha_3'})
\end{equation}
to separate the transformations contained in $Q_{\nu}$ from the $X_{\nu}^{\text{Diag}}$ in Eq.~\eqref{eq:Xdiag}.  From these relations, it is straightforward to obtain~\cite{bottomup}
\begin{equation}\label{eq:X0Qnu}
X_{\nu}^{\text{Diag}}=Q_{\nu}\times X_0^{\text{Diag}}.
\end{equation}
In analogy with Eq.~\eqref{eq:symdtond}, we then have\cite{bottomup}
\begin{equation}\label{eq:XUnu}
X_{\nu}=U_{\nu}X_{\nu}^{\text{Diag}}U_{\nu}^T,
\end{equation}
which relates $X_{\nu}^{\text{Diag}}$ to its generally non-diagonal form $X_{\nu}$.

For the charged lepton sector, we proceed as before from Eq.~\eqref{eq:defYe}.  We now have
\begin{equation}\label{eq:diagclCP}
(Y_{e}^{\text{Diag}})^{\dagger}M_e^{\text{Diag}}Y_e^{\text{Diag}}=(M_e^{\text{Diag}})^*=M_e^{\text{Diag}},
\end{equation}
in which we recall that $M_e$ is a Hermitian matrix.  From the form of Eq.~\eqref{eq:diagMe}, we see that $Y_e^{\text{Diag}}$ takes the form
\begin{equation}\label{eq:Ydiag}
Y_e^{\text{Diag}}=\text{Diag}(e^{i\gamma_1},e^{i\gamma_2},e^{i\gamma_3}),
\end{equation}
where $\gamma_i\in [0,2\pi)$, cf.~Eq.~\eqref{eq:Xdiag}. Hence, it should be clear that if we wish to separate the infinitely many nontrivial elements of $Q_e$ from $Y_e^{\text{Diag}}$, it is useful to define
\begin{equation}\label{eq:Y0}
Y_0^{\text{Diag}}=\text{Diag}(e^{i\beta'_1},e^{i\beta'_2},e^{i\beta'_3}),
\end{equation}
in which $\beta'_i=\gamma_i-\beta_i$ and $\beta_i$ is as defined in Eq.~\eqref{eq:baseunphyscl},  so that in analogy with Eq.~\eqref{eq:X0Qnu}, 
\begin{equation}\label{eq:Y0Qe}
Y_e^{\text{Diag}}=Q_e\times Y_0^{\text{Diag}}.
\end{equation}
It is interesting to note \cite{Girardi:2015rwa} that $Y_e^{\text{Diag}}$ can be related to its non-diagonal form with a relationship similar to that of Eq.~\eqref{eq:XUnu}, as follows:
\begin{equation}\label{eq:YUe}
Y_e=U_eY_e^{\text{Diag}}U_e^T.
\end{equation}
We note that even though both Eq.~\eqref{eq:XUnu} and Eq.~\eqref{eq:YUe} preserve the relationships in Eq.~\eqref{eq:defXnu} and Eq.~\eqref{eq:defYe} when $U_{\nu}\rightarrow U_{\nu}Q_{\nu}$ and $U_e\rightarrow U_eQ_e$, 
they are not generally preserved individually.  To see this explicitly,  let $Q_{\nu}'$ and $Q_e'$ represent a different choice of unphysical rephasings.  Applying $U_{\nu}\rightarrow U_{\nu}Q_{\nu}'$ and $U_e\rightarrow U_eQ_e'$ to Eq.~\eqref{eq:XUnu} and Eq.~\eqref{eq:YUe} yields
\begin{equation}\label{eq:X'Y'}
X_{\nu}\rightarrow X_{\nu}'=U_{\nu}Q_{\nu}Q_{\nu}'^2U_{\nu}^T\text{ and }Y_e\rightarrow Y_e'=U_eQ_eQ_e'^2U_e^T.
\end{equation}
 Since $Q_{\nu}'^2=1$ for any choice of phases in $Q_{\nu}'$, cf.~Eq.~\eqref{eq:baseunphysnu}, clearly $X_{\nu}'=X_{\nu}$ always.  On the other hand, $Q_e'^2\neq1$ for arbitrary choices of phases in $Q_e'$, cf.~Eq.~\eqref{eq:baseunphyscl}.   Therefore,  we generally have that $Y_e\neq Y_e'$.

This may naively seem problematic, but in actuality these results are consistent.  The consistency of the two mappings in Eq.~\eqref{eq:X'Y'} can be seen by observing that both  $Q_{\nu}Q_{\nu}'^2$ and $Q_eQ_e'^2$ are actually elements of the original set of unphysical phase choices.  In other words, $Q_{\nu}'^2=1$ is one of the possible choices of unphysical phase choices (albeit trivial) in Eq.~\eqref{eq:baseunphysnu}, and $Q_e'^2$ is one of the possible phase choices in Eq.~\eqref{eq:baseunphyscl}.  Therefore, both mappings in Eq.~\eqref{eq:X'Y'} reduce to those given in Eq.~\eqref{eq:XUnu} and Eq.~\eqref{eq:YUe} as long as we demand all such phase choices to exist.
However, it is important to note that it even though the mappings in Eq.~\eqref{eq:X'Y'}  can be defined consistently, they can in some cases have important implications for the preserved underlying symmetries.  

For example, such mappings may affect the explicit forms of the symmetry elements.  
In the charged lepton sector, the mapping $U_e\rightarrow U_eQ_e$  can fundamentally alter Eq.~\eqref{eq:YUe} because it (potentially) alters the trace of $Y_e^{\text{Diag}}$, as follows:
\begin{equation}\label{eq:TrYredef}
\text{Tr}(Y_e^{\text{Diag}})\rightarrow\text{Tr}(Q_eY_e^{\text{Diag}}Q_e^T)=\text{Tr}(Y_e'{}^{\text{Diag}}),
\end{equation}
and thus it can change the trace of $Y^{\text{Diag}}_e$ to that of $Y'_e{}^{\text{Diag}}$ if $Q_e^2\neq 1$.   As one may guess, the analogous  mapping in the neutrino sector, i.e., $U_{\nu}\rightarrow U_{\nu}Q_{\nu}$, does not have the same effect because it leaves Eq.~\eqref{eq:X'Y'} invariant, since $Q_{\nu}^2=1$.   Such a \textit{trace-changing} result for the $X_{\nu}^{\text{Diag}}$ can be produced through slightly different means, namely that of a global phase redefinition, cf.~Eq.~\eqref{eq:diagMnuphase}.  Such a global phase redefinition changes the traces of the $X_{\nu}^{\text{Diag}}$
thereby  affecting their group character when $\theta_{\nu}\neq 0$, cf.~Eq.~\eqref{eq:Xdiag}.  Thus, here we see another difference between the two sectors. Note also that Eq.~\eqref{eq:TrYredef} represents a transformation between the diagonal elements and will not transform to an object outside of the set $Q_e$ possibly \textit{unless} the original infinite set of symmetries has been reduced to a smaller more discrete set.  However, globally rephasing $M_{\nu}$ represents  an unphysical phase shift to the complex neutrino mases in Eq.~\eqref{eq:diagMnu}, and even by including all such symmetries/phase choices, may lead to something that can no longer be realized in a specific family symmetry group.




To this end, let us now turn to the consideration of generalized CP within the context of discrete lepton family symmetries.  It is very well known (see e.g.~\cite{Feruglio:2012cw,Nishi:2013jqa,bottomup,Dingnew,Chen:2015nha}) that a flavor symmetry element can be formed from the action of two generalized CP transformations.  For the neutrino sector, we see from  Eqs.~\eqref{eq:diagMnu} and~\eqref{eq:defXnu} that for the generalized CP elements $(X_\nu)_{i=0,1,2,3}\equiv X_i$, we have
\begin{equation}\label{eq:XtoG}
X_j^{\dagger}X_i^TM_{\nu}X_iX_j^*=M_{\nu},
\end{equation}
for $i,j=0,1,2,3$.  By comparing Eq.~(\ref{eq:XtoG}) to Eqs.~\eqref{eq:Xdiag}-\eqref{eq:X0Qnu}, it should be clear that the indices in the above are fixed by the Klein symmetry group as given in Eq.~\eqref{eq:diagGs}. Furthermore, we recall that for the phenomenologically desirable case of nondegenerate neutrino masses, $X_iX_j^*\in Z_2\times Z_2$.  Therefore, a flavor symmetry element $(G_\nu)_k\equiv G_k$ can be expressed as
\begin{equation}\label{eq:XX*toG}
G_k=X_iX_j^*
\end{equation}
for $i,j,k=0,1,2,3$.  For the charged leptons, in analogy, we find from Eqs.~\eqref{eq:diagMe} and~\eqref{eq:defYe} that the generalized CP elements $(Y_e)_{k,l}\equiv Y_{k,l}$
\begin{equation}\label{eq:YtoT}
Y_l^TY_k^{\dagger}M_eY_kY_l^*=M_e,
\end{equation}
which in turn implies that the flavor symmetry element $(T_e)_m\equiv T_m$ is given by
\begin{equation}\label{eq:YY*toT}
T_m=Y_kY_l^*,
\end{equation}
for some $m,k,l=0,\ldots,n-1$ if $G_e\cong Z_n$ or $m,k,l=0,\ldots, n_2+n_3-1$ if $G_e\cong Z_{n_2}\times Z_{n_3}$.\footnote{Note that we could not just apply one general CP transformation to $M_e$ to get a relationship between $T_e$ and $Y_e$ because the resulting space-time indices of the fields would be inconsistent.}  Here we note that it is important to realize that the phases $\alpha_i'$ contained in $X_0^{\text{Diag}}$, cf.~Eq.~\eqref{eq:X0}, are analogous to the $\beta_i'$ phases contained in $Y_0^{\text{Diag}}$, cf.~Eq.~\eqref{eq:Y0}.
Thus, Eq.~\eqref{eq:XtoG} and Eq.~\eqref{eq:YY*toT} demonstrate that it is possible to relate two successive generalized CP transformations to a flavor symmetry transformation.  Note that the complex conjugation in Eq.~\eqref{eq:XX*toG} and Eq.~\eqref{eq:YY*toT} plays a crucial role in that it allows the use of unitarity to reduce the products of flavor symmetry elements when expressed in terms of generalized CP symmetry transformations.

 The fact that family symmetry transformations can be expressed in terms of generalized CP transformations has further implications. As seen in Eq.~\eqref{eq:X0Qnu} and Eq.~\eqref{eq:Y0Qe}, it is also possible to express the generalized CP transformations in terms of flavor symmetry transformations.    By further assuming $Y_e$ and $X_{\nu}$ are elements of a discrete family symmetry group so that the set of possible phases $Q_e/Y_e$ can possess is reduced, there will exist integers $p$ and $q$ such that $(X_{\nu}^{\text{Diag}})^p=(Y_e^{\text{Diag}})^q=1$.  From Eq.~\eqref{eq:X0Qnu} and Eq.~\eqref{eq:Y0Qe}, we then have
\begin{equation}\label{eq:prodX}
(X^{\text{Diag}}_{\nu})^p=Q_{\nu}^p\times\text{Diag}(e^{ip\alpha'_1},e^{ip\alpha'_2},e^{ip\alpha'_3})=1
\end{equation}
and
\begin{equation}\label{eq:prodY}
 \begin{aligned}
&(Y^{\text{Diag}}_e)^q=Q_e^q\times\text{Diag}(e^{iq\beta'_1},e^{iq\beta'_2},e^{iq\beta'_3})=1.
\end{aligned}
\end{equation}
For the case of the charged lepton sector, we see that if Eq.~\eqref{eq:prodY} is ever to be satisfied without tuning $\beta_i$ contained in $Q_e$ and $\beta_i'$ against each other so that they cancel,
 $q$ must be a multiple of the order of $Q_e$ (such that $Q_e^q=1$) and $\beta'_i=2\pi k'_i/q$ for some $k'_i=0,1,\ldots, q-1$ (such that ($Y_0^{\text{Diag}})^q=1$).\footnote{In the unlikely case that $\beta_i+\beta'_i=0$ for  every $i$, then the orders of $Y_e$ and $T_e$ are identical.}
Following the same logic for the neutrino sector, it is straightforward to deduce that $\alpha'_i=2\pi a_i/p$ for some $a_i=0,1,\ldots, p-1$, in which $p$ must be an even integer ($Q_{\nu}^2=1$) to satisfy Eq.~\eqref{eq:prodX}. As a result, $X_{\nu}^{\text{Diag}}$ must be of \textit{even} order, confirming the results of \cite{Nishi:2013jqa}.  Additionally, $X_0^{\text{Diag}}$ from Eq.~\eqref{eq:X0Qnu} must  be the same even order (even though $G_0$ is order one) because $\alpha_i'=2\pi a_i/p$ for some $a_i=0,1,\ldots, p-1$,  which it has inherited from the nontrivial Klein group elements.    A similar situation occurs for $Y_0^{\text{Diag}}$ in Eq.~\eqref{eq:Y0Qe}; i.e., if $Q_e=1$, the order of the corresponding $Y_e^{\text{Diag}}$ need not be unity.

We note that the previous result that $p$ must be even and $q$ must be a multiple of the order of $Q_e$ may not hold in the nondiagonal basis, i.e.,  for $X_{\nu}$ and $Y_e$.   This cetainly is the case  if $U_{\nu}$ and $U_e$ are real so that $U_{\nu}=U_{\nu}^*$ and $U_e=U_e^*$, i.e., $U_eU_e^T=1$ and $U_{\nu}U_{\nu}^T=1$.  However, this may not always be true.  Thus with an eye towards deriving the most general conditions which $X_{\nu}$ and $Y_e$ must satisfy so that their orders are $p$ and $q$ respectively, we proceed by  inverting the transformations in Eq.~\eqref{eq:XUnu} and Eq.~\eqref{eq:YUe}:
\begin{equation}\label{eq:XYdiagpq}
(X_{\nu}^{\text{Diag}})^p=(U_{\nu}^{\dagger}X_{\nu}U_{\nu}^*)^p=1
\text{ and }
(Y_e^{\text{Diag}})^q=(U_e^{\dagger}Y_eU_e^*)^q=1.
\end{equation}
Taking the determinant of both of the above relations leads to
\begin{equation}\label{eq:detcondXY}
\text{Det}(U_{\nu}^*)^{2p}\, \text{Det}(X_{\nu})^p=1\text{ and } \text{Det}(U_{e}^*)^{2q}\, \text{Det}(Y_e)^q=1.
\end{equation}
We also note that the above conditions relating the determinants of $X_{\nu}$ and $Y_e$ to $U_{\nu}^*$ and $U_e^*$ are invariant under the transformation $U_{\nu}\rightarrow U_{\nu}Q_{\nu}$ and $U_e\rightarrow U_eQ_e$ because $Q_e^{2q}=Q_{\nu}^{2p}=1$.



\section{Generalization of Bottom-Up Constructions\label{sec:examples}}

We now discuss how these considerations allow for a generalization of the bottom-up approach given in \cite{bottomup}. This approach, which was based on the hypothesis that the full Klein symmetry is preserved in the neutrino sector such that lepton mixing is fully determined (up to charged lepton rephasings), is summarized as follows.   The Klein generators and the generalized CP symmetry elements of the neutrino sector can be expressed as a function of the lepton mixing parameters in the basis in which the charged lepton sector is diagonal.  Up to leptonic rephasings, in this basis $U_\nu$ can be written as
\begin{equation}
U_\nu = U^\text{Diag}_e U_\text{MNSP}.
\label{eq:unudiague1}
\end{equation}
Hence, for $U^\text{Diag}_e$ as the identity, $U_\nu$ can be parametrized in terms of the MNSP mixing parameters.  With this form of $U_\nu$,  
the Klein symmetry elements $(G_\nu)_i\equiv G_i$ and the generalized CP elements $(X_\nu)_i\equiv X_i$ can then be constructed explicitly as a function of the lepton mixing parameters, as follows\footnote{Recall that the $G_i$ are what we previously called $S_\nu$ in Eq.~(\ref{eq:symdtond}) with positive determinant.} (see Eq.~(\ref{eq:XUnu})):
\begin{equation}
G_i = U_\nu G_i^{\rm Diag} U_\nu^\dagger, \qquad X_i = U_\nu X_i^{\rm Diag} U_\nu^T.
\label{eq:symmelementsboth}
\end{equation}
This analysis was then carried out for several popular model scenarios for the MNSP mixing parameters \cite{bottomup}.  We note that in addition to working in the diagonal charged lepton sector basis, we also made simplifying assumptions in \cite{bottomup} about the leptonic sector phases.  More precisely,  we parametrized $U_\nu$ as follows:
\begin{equation}
U_\nu = P R_1(\theta_{23})R^\prime_2(\theta_{13},\delta) R_3(\theta_{12}),
\label{eq:unupaper1}
\end{equation}
in which the $R_i$ ($R_i^\prime$) are the usual (complexified) rotation matrices involving the leptonic mixing angles ($\theta_{12}$, $\theta_{23}$, $\theta_{13}$) and the leptonic Dirac phase $\delta$,  while $P$ is a diagonal matrix of the form $P={\rm Diag}(1,1,-1)$. This choice is thus very similar to the standard PDG form, except for the inclusion of $P$ and the neglect of the Majorana phases, which are included in the complex neutrino mass eigenvalues, as given in Eq.~(\ref{eq:diagMnu}). 

We will now revisit this analysis in the context of this work.  We will primarily focus once again on the case of a diagonal charged lepton sector, but now allowing for general leptonic phases.  As we will see, in this case useful relations can be obtained that can be of utility in top-down model building scenarios based on discrete groups.  We will also comment briefly on the case of a non-diagonal charged lepton sector, in which case the unphysical phases can play a role in the connection of the Klein and generalized CP symmetry elements in both sectors, depending on how $U_e$ and $U_\nu$ are reconstructed from the lepton data.

Starting with the case of the diagonal charged lepton basis, it is straightforward to see that the inclusion of nontrivial leptonic phases can affect the reconstruction of $U_\nu$ as given in Eq.~(\ref{eq:unudiague1}).  More precisely, since we have set $U_e$ to be the identity, the inclusion of a general phase matrix $Q_e$ means that Eq.~(\ref{eq:unudiague1}) is shifted to 
\begin{equation}
U_\nu = U_e^\text{Diag} Q_e U_\text{MNSP}.
\label{eq:QetoUnu}
\end{equation}
As a result, the phases in $Q_e$ can explicitly enter the Klein and generalized CP symmetry elements. Indeed, from Eq.~(\ref{eq:unupaper1}), the matrix $P$ itself can be interpreted as a specific choice of $Q_e$ (one with a negative determinant). As discussed in \cite{bottomup}, this choice was made for convenience in making an identification between the Klein and generalized CP symmetry elements with standard representations of elements of specific discrete groups.  However, another interpretation of $P$ can be understood from considering general leptonic phases, as follows.  Let us now take the case in which $U_\nu$ is instead parametrized by 
\begin{equation}
U_\nu = Q_e P R_1(\theta_{23})R^\prime_2(\theta_{13},\delta) R_3(\theta_{12}) \equiv P^\prime R_1(\theta_{23})R^\prime_2(\theta_{13},\delta) R_3(\theta_{12}),
\label{eq:rephasingparamue}
\end{equation}
where $Q_e$ are the charged lepton phases as given in Eq.~(\ref{eq:baseunphyscl}), and we define the matrix $P^\prime$ as 
\begin{equation}
P'= \text{Diag}(e^{i\phi_1}, e^{i \phi_2}, e^{i \phi_3}) \equiv Q_e P = \text{Diag}(e^{i\beta_1}, e^{i \beta_2}, -e^{i \beta_3}).
\label{eq:pprimedef}
\end{equation}
Therefore, in the basis in which the charged leptons are diagonal, the charged lepton rephasing degrees of freedom as discussed in this paper can be interpreted, in this context, as the following transformation of $U_\nu$:
\begin{equation}
U_\nu \rightarrow Q_e U_\nu.
\label{eq:newtmn}
\end{equation}
In direct contrast to the Klein symmetry transformations that we previously discussed, for which $U_\nu\rightarrow U_\nu Q_\nu$, it is clear that Eq.~(\ref{eq:newtmn}) is not a symmetry of Eq.~(\ref{eq:diagMnu}), but rather changes the specific $M_\nu$ that results in a given $M_\nu^\text{Diag}$. As a result, the Klein and generalized CP elements are necessarily modified, according to Eq.~(\ref{eq:symmelementsboth}).  As we will now discuss, these modifications can elucidate certain aspects of connecting these symmetry elements to specific elements of an assumed discrete symmetry group. 


 For the Klein generators, since the  $G_i$ are related to their diagonal forms via a unitary transformation (see Eq.~(\ref{eq:symmelementsboth})),  it is straightforward to see that the elements of $G_i$ are then modified from their forms as given in \cite{bottomup}, as follows:
\begin{equation}
(G_i)_{rs} \rightarrow (G_i)_{rs}\, e^{i(\beta_r-\beta_s)}, \qquad (r,s=1,2,3),
 \end{equation} 
 and we recall that $\beta_{1,2}=\phi_{1,2}$ and $\beta_3=\phi_3\pm\pi$ (see Eq.~(\ref{eq:pprimedef})). We see that the diagonal entries of the $G_i$ (and hence the trace) are unaffected by this rephasing, but the off-diagonal entries are changed.  In addition, the modified $G_i$ clearly satisfy the standard Klein relations
 \begin{equation}
 G_i^2=1, \qquad G_0 G_{i=1,2,3} = G_{i=1,2,3}, \qquad G_i G_j = G_k (i\neq j \neq k \neq 0).
 \end{equation}
 Such rephasings, while unphysical, can be helpful in the context of top-down model building based on discrete symmetry groups.  
 As an  example, there is a known connection in this context between the eigenvector of each of the $G_{i=1,2,3}$  with a positive $+1$ eigenvalue and the corresponding ($i$th) column of the MNSP matrix (up to permutations).  This one-to-one correspondence holds irrespective of whether nontrivial charged lepton rephasings are included in the parameterization of $U_\nu$ or not.   To see the ways in which including the phases can be informative, for concreteness let us express the Klein element $G_3$ as a function of the mixing parameters and the lepton rephasings.  Using Eq.~(\ref{eq:rephasingparamue}) and Eq.~(\ref{eq:pprimedef}), $G_3$ then takes the following form (see also \cite{bottomup}):
\begin{equation}
G_3 = \left (\begin{array}{ccc} -c^\prime_{13}&  e^{-i(\delta-\phi_1+\phi_2)}s^\prime_{13}s_{23} & e^{-i(\delta-\phi_1+\phi_3)}s^\prime_{13}c_{23}\\ e^{i(\delta-\phi_1+\phi_2)}s^\prime_{13}s_{23} & -c_{13}^2 c^\prime_{23}-s_{13}^2 & e^{i(\phi_2-\phi_3)} c_{13}^2 s^\prime_{23} \\  e^{i(\delta-\phi_1+\phi_3)} c_{23}s^\prime_{13} & e^{-i(\phi_2-\phi_3)}c_{13}^2s^\prime_{23} & c_{13}^2c^\prime_{23}-s_{13}^2 
\end{array} \right ),
\label{eq:g3gen}
\end{equation}
in which (as in \cite{bottomup}), $s_{ij}=\sin\theta_{ij}$, $c_{ij}=\cos\theta_{ij}$, $s^\prime_{ij}=\sin2\theta_{ij}$, and $c^\prime_{ij}=\cos2\theta_{ij}$.  

Let us now consider the class of models in which $\theta_{23}=\pi/4$ and $\theta_{13}=0$.  This includes the well-known tribimaximal (TBM) mixing scenario \cite{tbm}, for which the solar mixing angle is given by
\begin{equation}\label{eq:TBMang}
\theta_{12}^{\text{TBM}}=\tan^{-1} \left (\frac{1}{\sqrt{2}} \right ).
\end{equation}
It also includes other situations, such as golden ratio (GR1) mixing \cite{Datta:2003qg,Kajiyama:2007gx,GR1A5,minimalA5} for which 
\begin{equation}
\theta_{12}^\text{GR1} =  \tan^{-1}\left (\frac{1}{\phi}\right ),
\end{equation}
which depends on the golden ratio $\phi=(1+\sqrt{5})/2$, as well as other scenarios (see e.g.\cite{kingluhnreview} for a detailed review).
In this class of models, it is straightforward to see from Eq.~(\ref{eq:g3gen}) that the Klein element $G_3$ takes the form
\begin{equation}
G_3 = \left (\begin{array}{ccc} -1 &0& 0 \\ 0&0& e^{i(\phi_2-\phi_3)} \\ 0 & e^{-i(\phi_2-\phi_3)} &0 \end{array} \right ).
\label{eq:g3zero13max23}
\end{equation}
In the case of tribimaximal mixing, this can be identified with the canonical $U$ generator of the discrete group $S_4$, which is known to be the minimal group that contains the three Klein elements $SU$, $S$, and $U$ that generate tribimaximal mixing when they are preserved \cite{minimalS4}.  As is well known, in the group representation typically used in the literature (see e.g.\cite{S4CP}), the $S_4$ $U$ generator is given by
\begin{equation}
U = \left (\begin{array}{ccc} -1 & 0 &0 \\ 0& 0& -1 \\ 0&-1 & 0 \end{array} \right ).
\end{equation} 
Hence, for the choice of phases with $\phi_2-\phi_3=\pm \pi$, we see from Eq.~(\ref{eq:g3zero13max23}) that we can make the identification that  $G_3=U$.  Clearly, in addition, to make the connection with discrete groups, the phases $\phi_i$ need to be consistent with a specific subgroup of the discrete symmetry.  A minimal implementation of this condition is simply to set $\phi_2=0$ and $\phi_3=\pm \pi$, which results in $P'=P$, as used in \cite{bottomup}.   Similar statements can be made for the identification of $G_3$ as an element of $A_5$, the minimal group that results in GR1 mixing \cite{minimalA5}. The same considerations can be explored for the $G_{1,2}$ Klein elements, which also depend on $\theta_{12}$. It can easily be shown that the same phase choice results in the identification of $G_1=SU$ and $G_2=S$ of $S_4$ for the case of tribimaximal mixing, and that a similar identification holds for the elements $G_{1,2}$ in $A_5$ for GR1 mixing.

 An example in which $\theta_{13}\neq 0$ is the case of bitrimaximal (BTM) mixing \cite{BTmixing,BTmixingnoZn}, for which the neutrino sector mixing matrix  is given by
\begin{align}\label{eq:BTMmixingmat}
U^{\text{BTM}}_{\nu}&=\left(
\begin{matrix}
 \frac{1}{6} \left(3+\sqrt{3}\right) & \frac{1}{\sqrt{3}} & \frac{1}{6} \left(3-\sqrt{3}\right) \\
 -\frac{1}{\sqrt{3}} & \frac{1}{\sqrt{3}} & \frac{1}{\sqrt{3}} \\
 \frac{1}{6} \left(-3+\sqrt{3}\right) & \frac{1}{\sqrt{3}} & \frac{1}{6} \left(-3-\sqrt{3}\right) \\
\end{matrix}
\right).
\end{align}
For a diagonal charged lepton sector, the lepton mixing angles and Dirac CP-violating phase in this scheme are given by
\begin{equation}\label{eq:BTMang}
\theta_{12}^{\text{BTM}}=\theta_{23}^{\text{BTM}}=\tan^{-1}(\sqrt{3}-1),~\theta_{13}^{\text{BTM}}=\sin^{-1}\left (\frac{1}{6}(3-\sqrt{3})\right ), ~\delta^{\text{BTM}}=0.
\end{equation}
The BTM mixing pattern as outlined above  can naturally arise from the spontaneous breaking of a  $\Delta(96)$ flavor symmetry\cite{BTmixing,BTmixingnoZn}, which is in fact the smallest group for realizing BTM mixing.  Extending the discussion of \cite{bottomup} shows that with the inclusion of general charged lepton rephasings, we see for example that $G_3$ takes the form
\begin{equation}
G_3=\left (\begin{array}{ccc} -\frac{1}{3}-\frac{1}{\sqrt{3}} & \left (-\frac{1}{3}+\frac{1}{\sqrt{3}} \right )e^{i(\phi_1-\phi_2)} & \frac{1}{3} e^{i(\phi_1-\phi_3)} \\ \left (-\frac{1}{3}+\frac{1}{\sqrt{3}} \right )e^{-i(\phi_1-\phi_2)} & -\frac{1}{3} & \left (\frac{1}{3}+\frac{1}{\sqrt{3}} \right )e^{i(\phi_2-\phi_3)} \\ \frac{1}{3} e^{-i(\phi_1-\phi_3)} & \left (\frac{1}{3}+\frac{1}{\sqrt{3}} \right ) e^{-i(\phi_2-\phi_3)} & -\frac{1}{3}+\frac{1}{\sqrt{3}}  \end{array} \right ),
\end{equation}
which again reduces to the canonical form for $G_3$ as given in the literature for $\Delta(96)$ models in the case that $\phi_1-\phi_2=0$, $\phi_1-\phi_3=\pm \pi$, $\phi_2-\phi_3=\pm \pi$.  Again, a consistent and minimal implementation of this requirement is the choice of $P$, as before.


For the generalized CP symmetry elements, the situation is different since the $X_i$ are not related to their diagonal counterparts by a standard similarity transformation, but instead by $X_i = U_\nu X_i^\text{Diag} U_\nu^T$, as given in Eq.~(\ref{eq:XUnu}) and Eq.~(\ref{eq:symmelementsboth}).  We see that both the $Q_e$ phases and the overall Majorana phase shift $\theta_\nu$ (see the discussion just after Eq.~(\ref{eq:Xdiag})) affect both the diagonal and the off-diagonal entries of the $X_i$, and thus the traces are also affected. More explicitly, using Eq.~(\ref{eq:rephasingparamue}), the generalized CP symmetry elements given in \cite{bottomup} take the form
\begin{equation}
(X_i)_{rs} \rightarrow (X_i)_{rs}\, e^{i(\beta_r+\beta_s)}, \qquad (r,s=1,2,3),
\end{equation}
in which once again the $\alpha_i$ in the expressions for $X_i$ in \cite{bottomup} are to be replaced by $\alpha^\prime_i = \alpha_i+\theta_\nu$ (see Eq.~(\ref{eq:Xdiag})), and we recall the relation between the $\beta_i$ and $\phi_i$ as given in Eq.~(\ref{eq:pprimedef}).  As discussed in \cite{bottomup}, if the Majorana phases $\alpha^\prime_i$ are trivial, then for the choice of $\phi_1-\phi_2=0$, $\phi_1-\phi_3=\pm \pi$, $\phi_2-\phi_3=\pm \pi$,  for the case of tribimaximal mixing, the $X_i$ are identical to the Klein elements of the identity, $S$, $SU$, and $U$ of $S_4$, which are elements of the automorphism group of $S_4$.\footnote{Recall that a nontrivial prediction for the Dirac phase $\delta$ was obtained in $S_4$  for the case in which a single $Z_2$ flavor symmetry element was preserved \cite{S4CP}.}  Analogous statements can be made for the cases of GR1 mixing and bitrimaximal mixing in the case of trivial Majorana phases \cite{bottomup}.

However, in the case of bitrimaximal mixing and the connection to its minimal discrete group, $\Delta(96)$, there can also be nontrivial rephasings that yield new possible candidates for the generalized CP symmetry elements.  In such situations, we can be guided by the general relations given in Eq.~\eqref{eq:detcondXY}.  For the case in which the phases of $Q_e$ are set to zero, we have
\begin{equation}\label{eq:detcondXYBT}
\text{Det}(X_i^{\text{BTM}})^p=1,
\end{equation}
and hence the order of the product of the eigenvalues of $X_i^{\text{BTM}}$ must be an even integer $p$ (note that $\text{Det}(X_i)$ need not be 1, as this is not a necessary condition for Eq.~\eqref{eq:detcondXY}).   This can also be seen that by noting that for $Q_e=1$, $U_{\nu}^{\text{BTM}}=(U_{\nu}^{\text{BTM}})^*$, and hence Eq.~\eqref{eq:XUnu} is a similarity transformation that preserves the eigenvalues of $(X_i^{\text{BTM}})^{\text{Diag}}$. However, if nontrivial phases are included in $U_\nu$ as given in Eq.~(\ref{eq:rephasingparamue}), Eq.~(\ref{eq:detcondXYBT}) instead is modified to
\begin{equation}
\text{Det}(X_i^{\text{BTM}})^p=e^{2i(\phi_1+\phi_2+\phi_3)},
\end{equation}
and recall that when we consider discrete symmetry groups, the $\phi_i$ by necessity do not take on continuous values, but instead discrete values consistent with group transformations.

With these results in mind, we now consider the group theory of $\Delta(96)$, which is arguably all derivable from the character table of $\Delta(96)$, as given in~Table \ref{tab:Delta96-Char}.
\begin{table}[h!]
\centering
\begin{tabular}{|c|c|c|c|c|c|c|c|c|c|c|}
\hline
 $\Delta(96)$&$\bf{1}$ & $\bf{1^{\prime}}$ & $\bf{2}$ &$\bf{3}$ &$\bf{ \widetilde{3}}$&$\bf{\overline{3}}$ & $\bf{3^{\prime}}$&$\bf{ \widetilde{3}^{\prime}}$ & $\bf{\overline{3}^{\prime}}$ &$ \bf{6} $\\\hline
$\mathcal{I}$& $1$ & $1$ & $2$ & $3$ & $3$ & $3$ & $3$ & $3$ & $3$ & $6$ \\\hline
$3C_4$& $ 1$ & $1$ & $2$ & $-1+2 i$ & $-1$ & $-1-2 i$ & $-1+2 i$ & $-1$ &$ -1-2 i$ & $2$ \\\hline
$3C_2$&  $1$ & $1$ & $2$ & $-1$ & $3$ & $-1$ & $-1$ & $3$ & $-1$ & $-2$ \\\hline
$3C_4^{\prime}$&  $1$ & $1$ & $2$ & $-1-2 i$ & $-1$ & $-1+2 i$ & $-1-2 i$ & $-1$ & $-1+2 i$ & $2$ \\\hline
 $6 C_4^{\prime\prime}$ &$1$ & $1$ & $2$ & $1$ & $-1$ & $1$ & $1$ & $-1$ & $1$ & $-2$ \\\hline
 $32C_3$& $1$ & $1$ & $-1$ & $0$ & $0$ & $0$ & $0$ & $0$ & $0$ & $0$ \\\hline
$12C_2^{\prime}$&  $1$ & $-1$ & $0$ & $-1$ & $-1$ & $-1$ & $1$ & $1$ & $1$ & $0$ \\\hline
 $12C_8$ & $1$ & $-1$ & $0$ & $i$ & $1$ & $-i$ & $-i$ & $-1$ & $i$ & $0$ \\\hline
 $12C_4^{\prime\prime\prime}$ & $1$ & $-1$ & $0$ & $1$ & $-1$ & $1$ & $-1$ & $1$ & $-1$ & $0$ \\\hline
 $12C_8^{\prime}$ & $1$ & $-1$ & $0$ & $-i$ & $1$ & $i$ & $i$ & $-1$ & $-i$ & $0$\\\hline
\end{tabular}
\caption{The Character Table of $\Delta(96)$, where $kC_n$ denotes a conjugacy class of $k$ elements all of order $n$.\label{tab:Delta96-Char}}
\end{table}
The character table can be used to deduce the sum of the eigenvalues of $X_i^{\text{BTM}}$, as this is just the trace/character of this element.  Thus,  we can immediately restrict ourselves to considering the $3$-dimensional irreducible representations of $\Delta(96)$, i.e., the $\bf 3,3',\widetilde{3},\widetilde{3}', \bar{3},$ and  $\bf\bar{3}'$. We further restrict ourselves to the four faithful $3$-dimensional irreducible representations, i.e.,  the $\bf 3,3', \bar{3},$ and  $\bf\bar{3}'$, as the $\bf\widetilde{3}$ and $\bf\widetilde{3}'$ furnish unfaithful representations of $\Delta(96)$ that are isomorphic to $S_4$.     

For concreteness, let us first set the phases of $Q_e$ to zero, then consider the implications of a nonzero $Q_e$.  From Eq.~(\ref{eq:XUnu}), we thus obtain for a trivial $Q_e$ that
\begin{equation}\label{eq:Tr3delta96}
\text{Tr}(X_i^{\text{BTM}})=(-1)^{p_1} e^{i\alpha_1'} +(-1)^{p_2} e^{i\alpha_2'} +(-1)^{p_3} e^{i\alpha_3'},
\end{equation}
while allowing for nontrivial $Q_e$ yields
\begin{eqnarray}
\label{eq:generaltraceBTM}
\text{Tr}(X_i^\text{BTM})&=&\frac{1}{3}  \Big [(-1)^{p_1} e^{i\alpha_1'} \left (\rho_{11} e^{2i \phi_1}+ e^{2i \phi_2}+\rho_{13} e^{2i \phi_3} \right )  \\ &+& (-1)^{p_2}e^{i\alpha_2'}(e^{2i \phi_1} +e^{2i \phi_2}+e^{2i \phi_3}) +(-1)^{p_3}e^{i\alpha_3'} \left (\rho_{31} e^{2i \phi_1}+e^{2i \phi_2}+\rho_{33} e^{2i \phi_3}  \right ) \Big ],\nonumber 
\end{eqnarray}
in which 
\begin{equation}
\rho_{11}= \rho_{33}=1+\sqrt{3}/2, \qquad \rho_{13}=\rho_{31}= 1-\sqrt{3}/2.
\end{equation}

We now recall the discussion of the BTM example in~\cite{bottomup}, in which it was posited that for this case,  $\alpha_1=\alpha_3=\frac{\pi}{6}$ and $\alpha_2=-\frac{\pi}{3}$, so that $\alpha_{31}=\alpha_3-\alpha_1=0$ and $\alpha_{21}=\alpha_2-\alpha_1=-\pi/2$, to match known results in the literature\cite{Ding:2014ssa}.  Let us first consider the case in which the phases of $Q_e$ are trivial. With these assumptions, we can see that for $X_0^\text{BTM}$ and $X_2^\text{BTM}$, their traces satisfy 
 \begin{equation}\label{eq:unshiftXBT0}
\text{Tr}(X_{0}^{\text{BTM}})=e^{2\pi i/3+ i\theta_\nu}(-1-2i),
\end{equation}
and 
\begin{equation}\label{eq:unshiftXBT2}
\text{Tr}(X_{2}^{\text{BTM}})=e^{2\pi i/3+ i\theta_\nu}(-1+2i).
\end{equation}
We thus see that if we take the overall unphysical phase shift $\theta_{\nu}$ to be $\theta_{\nu}=4\pi/3+2n\pi$, then Eqs.~\eqref{eq:unshiftXBT0} and~\eqref{eq:unshiftXBT2} are in agreement with the entries for the $3C_4$ and $3C_4^\prime$ in Table \ref{tab:Delta96-Char}.  This phase shift must also be made so that the order of the $X_i^{\text{BTM}}$ is changed from $12$ to $4$ as $\Delta(96)$ does not have a $C_{12}$ conjugacy class, as seen from Table \ref{tab:Delta96-Char}.  Similar statements can be made for the $X^\text{BTM}_1$ and $X^\text{BTM}_3$ elements, with the identification of either the $6C_4^{\prime\prime}$ or the $12C_4^{\prime\prime\prime}$ conjugacy classes. We also note that in each case, the order of $X_i$ is a multiple of 2, as previously discussed.  Therefore, the order of the nontrivial $\Delta(96)$ Klein symmetry elements is consistent with this implementation of BTM mixing. 

For nontrivial $Q_e$, there is more freedom to match to specific discrete groups. Here it is not just the trace of $X_i$ that must be fixed, as it is also important to ensure that $X_i$ respects the group multiplication laws.  In certain cases, these conditions require that the $\phi_i$ take trivial values ($0$ or $\pm \pi$), i.e.,~the rephasing symmetry associated with the $\phi_i$ must be restricted to a $Z_2\times Z_2\times Z_2$ subgroup along with an overall phase that must itself be restricted by the group multiplication laws.  This can easily be seen from the form of Eq.~(\ref{eq:generaltraceBTM}), which does not distinguish between $\phi_i=0$ and $\phi_i=\pm \pi$. In the case previously discussed of connecting the BTM symmetry elements with group elements of $\Delta(96)$ with the nontrivial choice of $\alpha_i^\prime$ as given above, it is straightforward to show that the $\phi_i$ must indeed be constrained in this way to satisfy the group multiplication laws (with an overall phase of  $e^{\pm( i\pi/2+2n\pi)}$, which we note is consistent with order 4 symmetry elements).  With other choices of $\alpha_i^\prime$ and $\phi_i$, there may in principle be other connections of interest that can be made to different discrete groups for a given mixing angle pattern.

Up to this point, we have taken $U_e$ to be the identity, and in so doing, absorbed the effects of $Q_e$ in the parametrization of $U_\nu$ (see Eq.~(\ref{eq:rephasingparamue})).   It is of course also possible, and indeed must be physically equivalent, instead to keep $Q_e$ in the charged lepton sector.  In either case, the diagonal form of $U_e$ then also usually implies the existence of a residual, diagonal charged lepton flavor symmetry\cite{kingluhnreview} that contains elements that are a subset of the possibilities represented by $Q_e$ of~Eqs.~\eqref{eq:diagGe2gen}-\eqref{eq:diagGe1gen}.   As previously discussed, the unphysical phases contained in the residual charged lepton symmetry should all be distinct, or additional free parameters will in general arise, ``forcing'' the charged lepton mixing away from the identity unless these parameters are tuned accordingly.  Since we are working in the diagonal charged lepton basis, $M_e$ is real and diagonal. The invariance condition given in Eq.~\eqref{eq:detcondXY} with $U_e=1$ then implies that for the case at hand (neglecting $Q_e$), the generalized CP elements $Y_e$ obey the relation
\begin{equation}
\text{Det}(Y^{\text{BTM}}_e)^q=1,
\end{equation}
in which $q$ is a multiple of the order of the corresponding residual charged lepton flavor symmetry.  We also note that all possible generalized CP symmetry elements that are consistent with $U_e=1$ can be found from  Eq.~\eqref{eq:Y0Qe}.  

Finally, it is worthwhile to comment on the situation for which a general basis is chosen such that neither the charged leptons or the neutrinos are diagonal. The reason is that even though a basis change can always been made to diagonalize either sector, a general basis may facilitate the connection between the flavor and generalized CP symmetry elements and the representations of a specific discrete group.  In this case, within the bottom-up construction there is then a question of how the observed lepton mixing parameters are split between the two sectors, i.e.~the choice of $U_e$ and $U_\nu$ such that $U_e^\dagger U_\nu = U_\text{MNSP}$ (up to lepton rephasings).  When $U_e$ is not the identity, clearly the rephasing matrix $Q_e$ is not so easily translated into $U_\nu$, as in Eq.~(\ref{eq:QetoUnu}).  However, $U_e$ and $U_\nu$ can be transformed more generally, as follows:
\begin{equation}
U_e \rightarrow \widetilde{Q} U_e Q_e, \qquad U_\nu \rightarrow \widetilde{Q} U_\nu Q_\nu, 
\end{equation}
in which the transformation given by $\widetilde{Q}$ clearly leaves $U_\text{MNSP}$ invariant.  The matrix $\widetilde{Q}$ can be a full $U(3)$ transformation, or it can be a subset of this full set of transformations, such as a $U(1)^3$-preserving transformation.  The specific choice of $\widetilde{Q}$ thus also has obvious implications for the bottom-up construction of the symmetry elements, in analogy to the effects discussed here in the diagonal case.  Once again, it may be that such lines of reasoning open up new model-building directions in the context of family and generalized CP symmetry groups.

In summary, given that there can be a mismatch in the way in which the mixing angles of $U_\nu$ and/or $U_e$ are parametrized and the ways in which group representations of discrete groups are given in the literature, the unphysical leptonic phases can be of utility in connecting the bottom-up construction of symmetry elements to specific discrete group representations.  The case in which the charged leptons are taken to be diagonal is just one simple example.   That being said, since including $Q_e$ into $U_\nu$ is {\it a priori} not necessary since by definition it can be removed by rephasing the charged lepton fields, an equivalent alternative is that these group elements can be shifted via a unitary transformation such that they align with a trivial reconstruction of the Klein generators in this context, based on the standard parametrization of $U_\text{MSNP}$ and setting $P'$ as the identity.  A further (equivalent) alternative is to carry out the bottom-up construction with a different parametrization of $U_\text{MNSP}$, as clearly there is nothing sacred from the model-building point of view about the PDG parametrization.  However, selecting a specific parametrization and including these phases in the construction of the Klein generators and the generalized CP symmetry elements allows for this freedom to be taken into account in a straightforward way that can facilitate the identification of viable discrete groups for top-down flavor model building.


\section{Conclusions\label{sec:conclusion}}


If experiments reveal that neutrinos are Majorana particles, the possibility exists that there is a residual symmetry in the neutrino sector that completely fixes $U_{\text{MNSP}}$ in the diagonal charged lepton basis, up to rephasing by unphysical charged lepton phases.   However, such a symmetry cannot make predictions for Majorana phases.  In order to produce such predictions, a popular and well-motivated approach is to impose a generalized CP symmetry (consistently) alongside of the flavor symmetry and spontaneously break both symmetries (presumably at a high scale, such as the unification scale) to generate mixing angle and phase predictions, accordingly.  In such a top-down approach, the angle and phase predictions that arise from this breaking become subject to model-dependent corrections such as renormalization group evolution, canonical normalization and corrections from subleading contributions to either the charged lepton or neutrino sectors.\footnote{One may anticipate such corrections to be subleading because renormalization group and canonical normalization effects are expected to be small in realistic models with hierarchical neutrino masses, and the charged lepton corrections are typically at most Cabibbo-sized\cite{CN, genrge,CLRGECN}. It is possible to have large charged lepton corrections if $U_{\nu}$ is taken as the starting point for $U_{\text{MNSP}}$, relying solely on $U_e$ for corrections to bring $U_{\nu}$ to the experimentally measured values\cite{CL}.}  Alternatively, we can start from a bottom-up approach, in which the flavor and generalized CP symmetry elements can be constructed explicitly based on specific mixing angle scenarios, and thus can be used as a roadmap for top-down model building.  In this work, we have investigated the effects of considering nontrivial unphysical  lepton sector phases in this context, focusing on their group-theoretical properties.  We also have discussed how such lepton sector rephasings extend the results of \cite{bottomup}, in order to further elucidate the interplay between generalized CP and flavor symmetries in the charged lepton and neutrino sectors.

By extending the results of \cite{bottomup}, we have put forth a more complete bottom-up approach that incorporates nontrivial, unphysical charged lepton phases as well as unphysical shifts to Majorana phases.  Our analysis further identifies the similarities and differences between generalized CP symmetries in the charged lepton and neutrino sectors while further elucidating the group properties of the generalized CP symmetry elements.  The results provide a set of group theoretical relations that must be satisfied at low energies for all models within this general framework. To this end, the methods outlined here can serve as guidance for future model-building by further clarifying the effects that preserving various residual generalized CP and flavor symmetry elements can have on models of lepton masses and mixing angles.


\section*{Acknowledgements}


A.S. acknowledges support from the
research grant 2012CPPYP7 (Theoretical Astroparticle Physics) under the program PRIN
2012 funded by the Italian Ministry of Education, University and Research (MIUR) and support from
the ERC Advanced Grant No. 267985 Electroweak Symmetry
Breaking, Flavour and Dark Matter: One Solution for Three Mysteries (DaMeSyFla),
as well as partial support from the European Union FP7 ITN INVISIBLES (Marie Curie Actions,
PITN-GA-2011-289442-INVISIBLES).  L.E. is supported by the U.S.~Department of Energy under the contract DE-FG-02-95ER40896.  L.E. also thanks the Enrico Fermi Institute for its hospitality during the early stages of this work.


\end{document}